\documentclass{article} 
\usepackage{graphicx}

\usepackage{amsfonts}
\usepackage{amsmath}

\newcommand{\R}{\mathbb{R}}
\newcommand{\Z}{\mathbb{Z}}

\begin{document}
\title{Hierarchical Matching and Regression with Application to Photometric 
Redshift Estimation}

\author{Fionn Murtagh \\
email: {\tt fmurtagh@acm.org}}


\maketitle


\begin{abstract}
This work emphasizes that heterogeneity, diversity, discontinuity, and 
discreteness in data is to be exploited in classification and regression 
problems. A global a priori model may not be desirable.  For data analytics 
in cosmology, this is motivated by the variety of cosmological objects such 
as elliptical, spiral, active, and merging galaxies at a wide range of 
redshifts.  Our aim is matching and similarity-based analytics that takes 
account of discrete relationships in the data.  The information structure of 
the data is represented by a hierarchy or tree where the branch structure, 
rather than just the proximity, is important.  The representation is related 
to p-adic number theory.  The clustering or binning of the data values, 
related to the precision of the measurements, has a central role in this 
methodology.  If used for regression, our approach is a method of cluster-wise 
regression, generalizing nearest neighbour regression.  Both to exemplify this 
analytics approach, and to demonstrate computational benefits, we address the 
well-known photometric redshift or `photo-z' problem, seeking to match Sloan 
Digital Sky Survey (SDSS) spectroscopic and photometric redshifts. 
\end{abstract}


\section{General Introduction}

We are concerned with matching, and drawing inferences (extrapolation and interpolation, 
prediction, distributional degree of association, etc.) from structures that are 
discrete.  In addition to being discrete, there are associations, similarities and 
identities that are relevant.  Also relevant are incorporation, inclusion, properties
of an object being a subset of properties of one or more other objects.  

A reasonable and a natural representation for such structures is an ultrametric or tree 
topology.  Our objects are taken as nodes of a tree. A tree is a synonym for hierarchy. 
These objects, or entities, could, if desired, include sub-objects and sub-entities 
also.  In set notation, a hierarchy is a partially ordered set, or poset: let sets 
$q, q^\prime$, associated with nodes, have a least common parent node, 
$q^{\prime\prime}$.  Parent/child node ordering uses 
$q, q^\prime \subset q^{\prime\prime}$.  For a partially ordered set, it is required 
that for any two subsets. $q, q^\prime \in 2^I$, for power set $2^I$ where $I$ is 
the union of all sets, or the universal set: one class is a subset of the other, as 
the only permitted overlapping: $q \cap q^\prime = \emptyset$ or either 
$q \subset q^\prime$, or $q^\prime \subset q$.  

\section{Real Number System: Practicalities of p-Adic and m-Adic Number Systems}
\label{subsecrealadic}

In our Baire, longest common prefix, metric, from which we can directly (i.e.\ in 
linear computational time) read off a hierarchy, we must explain why the resulting 
clusters are meaningful in the following sense: why is it reasonable to have different 
clusters, from the top or first level or partition onwards, for 2.9999 and 3.000, for 
example?

Firstly, our approach assumes precise measurement to a given precision.
Our Baire method favours contexts where the digits of precision of
measurement are ordered (decreasing importance associated with increasing
digit of precision).  Hence our Baire methodology is designed for fast,
exact proximity matching.

So -- tree branch is what is important here, rather than just the proximity, alone, of 
singletons.  We might even say, if the singletons are derived from, or originate from, 
the root, then precisely how they came about from the root, i.e.\ their path from the 
root, that is what we want to especially take into consideration.   In this very 
particular sense then, we have that 0.50 is distinct from 0.49, and, we reason further, 
0.500 is distinct from 0.499, and so on, just due to the account taken of digit 
priority.  

A further perspective on this, and justification for the role of digit sequence, is as 
follows.  Consider that we are measuring with these two exemplary numbers here, and 
what we additionally want to do is to ``take time into account'', with time steps, say
milli-seconds, being used for determining each additional digit in what we are measuring.
We are determining our numbers, digit by digit.  In each such digit stage, or, 
informally expressed, in each such ``time-step'', we are adding detail and precision to 
our measured values.  

Of course it is to be accepted that in a real number system, by convention a p-adic 
number system when p $ = \infty$, 0.5 is identical to, and can be expressed as 
0.499999....  Irrespective of whether taking real numbers as an m-adic number system, 
with m = 10, or, with their important mathematical properties, considering p-adic number 
systems, for p prime, our concluding remark is as follows.   We take into consideration 
a priority order of digits.

One further example of this adic number representation is as follows.  In 
\cite{capillaries}, the dynamics of fluid flows in 
tree branchings is at issue.  This 
is with application to petroleum underground reservoirs.  

\section{Previous Work}

\subsection{Determining Photometric Redshifts from Colour and Magnitude 
Observed Data, and Evaluating relative to Spectroscopic Redshifts}

In \cite{vanzella} there is predicting of photometric 
redshifts ``from an 
ultra deep multicolor catalog''.  Training is carried out with spectroscopic
redshifts.  This is noted: ``the difficulty in obtaining spectroscopic 
redshifts of faint objects'', and then: ``A crucial test in all cases is the
comparison between the photometric and spectroscopic redshifts which is 
typically limited to a subsample of relatively bright objects''.  The 
SDSS DR1 catalogue used is ``almost entirely limited to $z < 0.4$''.  The 
test sample is 88108 galaxies, and 24892 galaxies in the training sample.  
It is noted how other approaches to nonlinear regression, including a Bayesian
method, polynomial regression and nearest-neighbour regression are claimed
to perform worse (citing \cite{csabai}).  The latter 
work, \cite{csabai}, 
uses approximately 35000 galaxies with spectroscopic redshifts, from the
SDSS EDR (Early Data Release) database, pre-DR1 (Data Release 1, the most 
recent at the time of writing, being DR-12).  In \cite{csabai}, 
$0.2 < z 
< 0.3$ redshifts are used. Colour and magnitude data are used to estimate
photometric redshifts.  In \cite{firth}, a training set 
of 10000 and a test
set of 7000 SDSS objects (galaxies but indicating that stars are also included,
with $z < 0.5$).  

\subsection{Interval Measurements for Bayesian ``stacking'' Modelling; 
Accuracy and Correctness of Measurement} 

In \cite{shu} velocity distributions are at issue, for association 
with galaxy 
sizes, to ``determine `dynamical masses' that are independent of stellar-population
assumptions'', with that to be used for evolution of galaxies for given mass, 
following relationship estimation with mass and gravitational potential.  Interest
is in elliptical galaxies, that are ``To a first approximation ... `pressure-supported' 
rather than rotationally supported''.  Velocity dispersion is to be based on 
spectroscopic data.  Now, in particular for faint, even if luminous, galaxies, 
there will be uncertainty and non-Gaussianity in measurement.  From SDSS III, 
430000 galaxies are used, primarily with redshifts $0.2 < z < 0.8$.  Eigenspectra
are determined from principal components analysis.  Because of the imprecision of
measurement the following is carried out, in the estimation of velocity dispersion.
Both in redshift and in absolute magnitude, respectively with intervals of 0.04 and 0.1,
galaxies are binned.  Therefore, for error or imprecision of measurement, binning,
i.e.\ interval measurements, are a way to somewhat robustify the data.  Based on 
extensive analyses, it is concluded that here the ``stacking'' of multiple spectra 
is replaced by a new ``Bayesian stacking'' approach.  (The hierarchical Bayesian 
approach is summarized in section 3.2 of \cite{shu}).

In \cite{bolton}, use is also made of
the work of \cite{shu}.  Under ``Known issues'', there are the 
following: the use
of probability priors on principal component analysis coefficient combinations; 
spectra that are obscured by others, e.g.\ quasar spectra, by AGN spectra; spectra
affected by ``cross-talk from bright stars''; superpositioning of observed objects; 
and a few classes of object, and detector suitability (``fibers near the edge of the 
spectrograph camera fields of view'').

\subsection{Nonlinear Regression}

In \cite{dabrusco}, multilayer perceptrons (MLPs) are 
used to relate photometric redshifts
to spectral information.  Varying object classes (normal galaxies, stars, late type
stars, nearby AGNs, distant AGNs) are subject to principal component analysis of 
spectra, to provide an eigenvector-based spectral classification index. 

The case is then made for carrying out the nonlinear regression, using MLP, on 
two different redshift intervals, $z < 0.25$ and $z > 0.25$.  Differing galaxy 
populations are associated with these redshift intervals.  A total of 449370 
galaxies were studied.   Just interestingly, consideration was given to not too 
full redshift intervals used for training, but $[0.01, 0.25] \subset [0.0, 0.27]$ 
and $[0.25, 0.48] \subset [0.23, 0.50]$.  

The foregoing work is pursued in \cite{dabrusco2}.  
SDSS DR4 data was used.   
A most comprehensive introduction is provided (section 1, Introduction, section 2, 
Photometric redshifts).  Included is the following note: ``photometric redshift 
samples are useful if the structure of the errors is well understood'', because 
this points to the reliability of, and confidence in, measurement.  Later there are
these statements: 
``photometric redshift estimates depend on the morphological type, age, metallicity, 
dust, etc. it has to be expected that if some morphological parameters are taken 
into account besides the magnitudes or colors alone, estimates of photometric
redshifts should become more accurate.''  In this work, the ``near universe'', 
$z < 0.5$, is at issue, and also with discussion of ``the near and intermediate 
redshift universe'', $z < 1$.  As before, two separate MLPs were used on this data,
for ``nearby'', $z < 0.25$, and for ``distant'', $z > 0.25$ objects.   These objects
comprised 449370 galaxies.  

A most revealing statement is the following: 
``the derivation of photometric redshifts requires, besides an accurate evaluation 
of the errors, also the identification of a homogeneous sample of objects.''

\section{Data Analysis}

\subsection{Data and Objectives To Be Pursued}

SDSS (Sloan Digital Sky Survey) data used was from Data Release 5, 
\cite{dr5}, 
relating to the following: ``Stripe 82 is an equatorial region repeatedly imaged 
during 2005, 2006, and 2007'', \cite{stripe82}.  Data were as follows: 
number of objects: 443094; right ascension, declination, spectroscopic redshift, 
photometric redshift.  Then minimum redshifts, respectively spectroscopic and
photometric, are: 0.000100049, 0.0001035912, and the maximum redshifts are: 
0.599886, 0.5961629.  

Our objective is to assess spectroscopic redshift from photometric redshift.  While 
regression, whether classical linear (statistical least squares), or nonlinear 
(multilayer perceptron, k-nearest neighbour, etc.) are relevant, we seek the 
following.  

\begin{enumerate} 
\item Take the discreteness of measurement into account. 
\item Therefore, we take distinction of value to be primarily associated with the 
discrete sourcing of our measurements, rather than being solely a statistical 
uncertainty or error component of our measurement. 
\item However statistical uncertainty or error component of measurement are taken as 
integral to the discreteness of sourced data.
\item It arises from this reasoning that what is important in practice is to be 
able to codify one's data, in the sense both of data encoding and of data representation,
here related to number theory.
\item From the data encoding and representation, we are seeking to associate data 
interpretation and understanding, with the discrete sourcing of our measured data.
\end{enumerate}

While cosmology presents the primary motivation for our m-adic and p-adic analytics, 
applications and opportunities for similar perspectives are numerous in other 
sciences also.  (A small set of notes follow.  Notationally $m > 2$ is typically a
positive integer, and $p$ is prime.  An m-adic number system is a ring, while a p-adic
number system constitutes a field.  A field has a multiplicative inverse for non-zero
values, i.e.\ it permits division.)  

\subsection{Preliminary Data Analysis}

\begin{figure}
\begin{center}
\includegraphics[width=6cm]{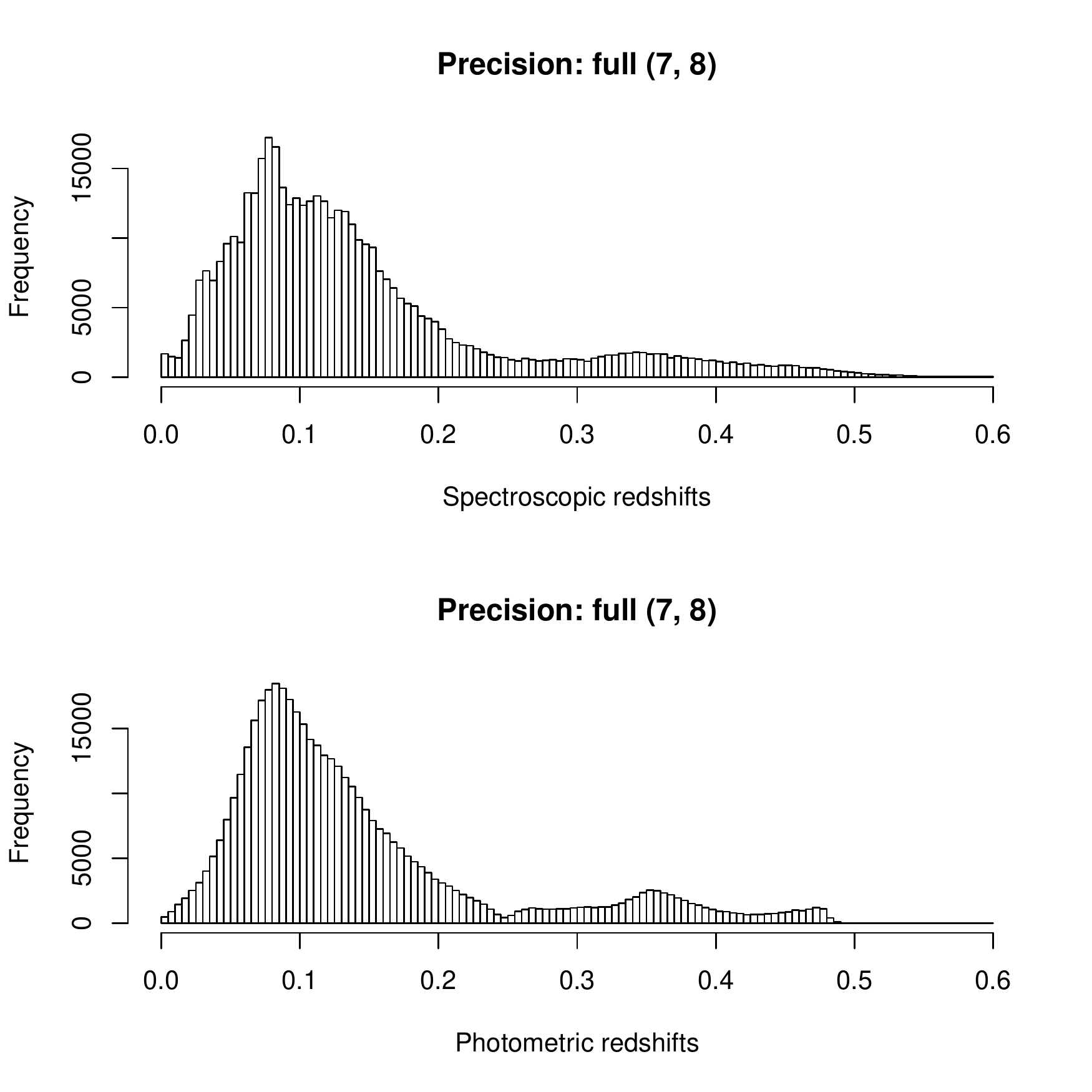}
\caption{Full precisions, 7 or 8 digits. Top, spectroscopic, and 
bottom, photometric.}
\label{fig1}
\end{center}
\end{figure}

For initial exploratory analysis purposes, we consider the histograms, cf.\ 
Figure \ref{fig1}. While mainly peaked
around the lower redshifts, there are some other interesting smaller peaks.  The 
digits of precision of the data are at issue.  It is to be noted that some data
values are limited to about three digits of precision.  


A first analysis will look at the very first digit of precision.  From the 
443094 objects, we find that a 0 first digit of precision is shared by 
162034 spectroscopic and photometric redshifts; a 1 first digit of precision is
common to 144602 redshifts; a 2 first digit of precision is common to 22643
redshifts; a 3 first digit of precision is common to 26441 redshifts; 
a 4 first digit of precision is common to 11166 redshifts; and a 5 first digit 
of precision is common to 21 redshifts.  Overall, the first digit of precision 
of the spectroscopic and photometric redshifts is common to 366907 objects, 
that is, 83\% of all cases.  

This is encouraging to begin with.  It indicates one relevant and useful way 
to determine commonality, or association, between the more reliable spectroscopic
redshifts and the possibly more accessible photometric redshifts.  The next stage
of our analysis is to see if this finding, of 83\% commonality of spectroscopic and 
photometric redshift measurement, can be furthered.  

If we look at both the shared first digit of precision, and additionally a difference 
in the first digit of precision of at most 1, then we find that 99.6\% of all the 
spectroscopic and photometric redshift measurement are that close in measurement value. 
While this is motivational, it requires further study of just what redshifts differ
by 1 in the first digit of precision.  However, we do not consider such a finding as 
generally and broadly applicable.  

\section{Re-Representing Our Data in p-Adic and Other Number Systems}

\subsection{Number Systems Other Than Real}

In the regression-oriented matching of spectroscopic redshift and photometric
redshift values, motivation for our approach is as follows. 

With reference to the histograms displayed in the previous section, in 3-dimensional 
Euclidean space, assumed distribution functions could be used to calibrate one 
such distribution function against another.  This ``calibration'' could be cluster-based,
through determining, for example, a Gaussian mixture fit to the assumed distribution 
functions.  Gaussian model-based mixtures can also be hierarchical, providing 
model-based cluster trees, \cite{clustertrees}. Another 
viewpoint could be to use 
RA and Dec local dependencies.  That could imply the regression of RA, Dec, 
z$_s$ on z$_p$ (the latter denoting spectroscopic redshift, photometric redshift).

Now, compared to a Euclidean and Hilbert space, we are dealing with discrete object
locations and clustered, albeit delimited, regions of objects.  A graph and more
particularly, a tree is an appropriate representation, rather than a continuous space. 
Although a side remark in the current context, it was noted in section 
\ref{subsecrealadic} how an ordered time dimension, in particular through being ordered
rather than being real-valued, can also be subsumed in this approach.  

Because of the directly mapped, rooted tree representation that can be associated with 
any m-adic number representation, we proceed as follows: consider our given decimal
or base 10 measurements, as m-adic with m = 10.  Efficiently derive other m-adic 
number representations, to assess them.  

\subsection{Re-Representing Data in Other Number Systems, through Efficient Approximation}

In \cite{sparsecoding}, the following innovative approach was 
developed for 
re-representing a data set, represented $m$-adically by a closest fit approximation
by a data set, represented $(m-1)$-adically.  

For all neighbour or adjacent digit values, at a given precision level, that have the 
same parent digit value, i.e.\ at these digit values' immediate preceding precision 
level, assess the following.  Firstly, if these neighbour values are identical, then 
there will be no intervention.  Secondly, if these neighbour values differ by more than 
1, then there will be no intervention.  Thirdly, if these neighbour values differ by 
1, then assessment is made of what overall pair of such values, with the same 
parent value, and differing by 1, are such that their cardinality is minimal.  We are 
going to merge this set of neighbour values.  The following properties of this 
processing are as follows.  By design, this constitutes a minimal overall change in 
our data.  This implies one digit less, in the entirety of data representation. Therefore
this is a best approximation to our data, starting from an m-adic representation, and 
passing to an m$-1$-adic representation.  There remains one final part of this 
processing: from the chosen set of neighbour digit values, the larger of these two 
values is altered to the smaller of these two values.  That is carried out in the 
data.  Very finally, for consistency and coherence of number representation, all values 
that were greater than this modified value are decreased by 1.  

The effectiveness of this approach was demonstrated in 
\cite{sparsecoding}.  
Computationally it is linear in the numbers of objects multiplied by the number of 
digits of precision.  That is, it is linear in the data set size, expressing the total 
number of digits.  

\subsection{m-Adic Fitting of Spectroscopic Redshifts}

As described in the previous subsection, we fit the given m-adic data, for $m = 10$, 
with the best fitting $m = 9$-adic representation; then with the (stepwise) 
best fitting $m = 8$-adic representation; then with the (stepwise) best fitting 
$p = 7$-adic representation;  then with the (stepwise) best fitting $m = 6$-adic 
representation; then with the (stepwise) best fitting $p = 5$-adic representation;
then with the (stepwise) best fitting $m = 4$-adic representation; then with the
(stepwise) best fitting $p = 3$-adic, or ternary, representation; and finally 
with the (stepwise) best fitting $p = 2$-adic, or binary, representation.

\begin{figure}
\begin{center}
\includegraphics[clip,trim={0 4cm 0 4cm},width=6cm]{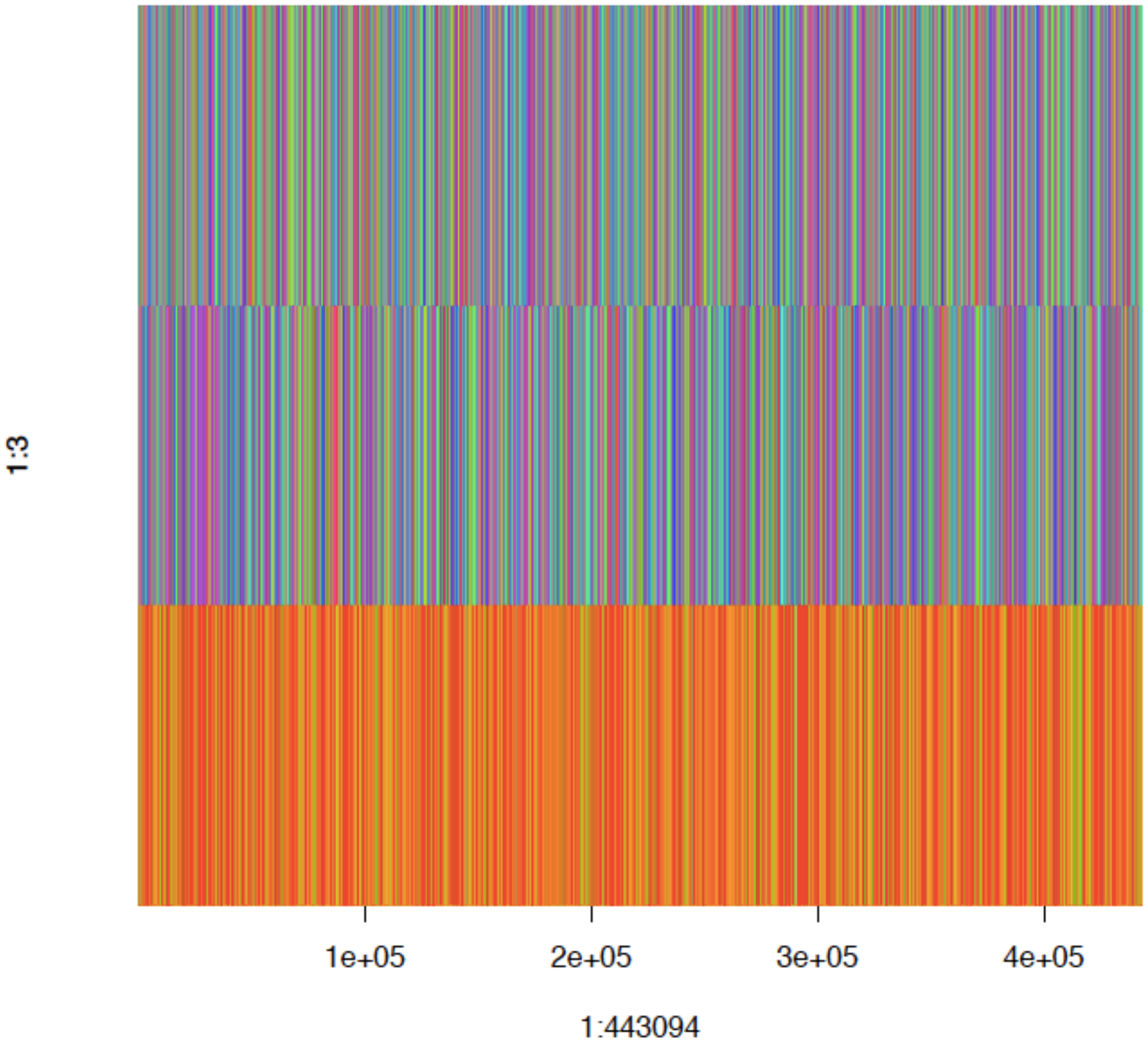}
\caption{Spectroscopic redshifts. 
Initial m-adic display, for m = 10.  Three digits of precision used, increasing
on the ordinate. The abscissa lists the spectra.}
\label{fig5}
\end{center}
\end{figure}

\begin{figure}
\begin{center}
\includegraphics[clip,trim={0 4cm 0 4cm},width=6cm]{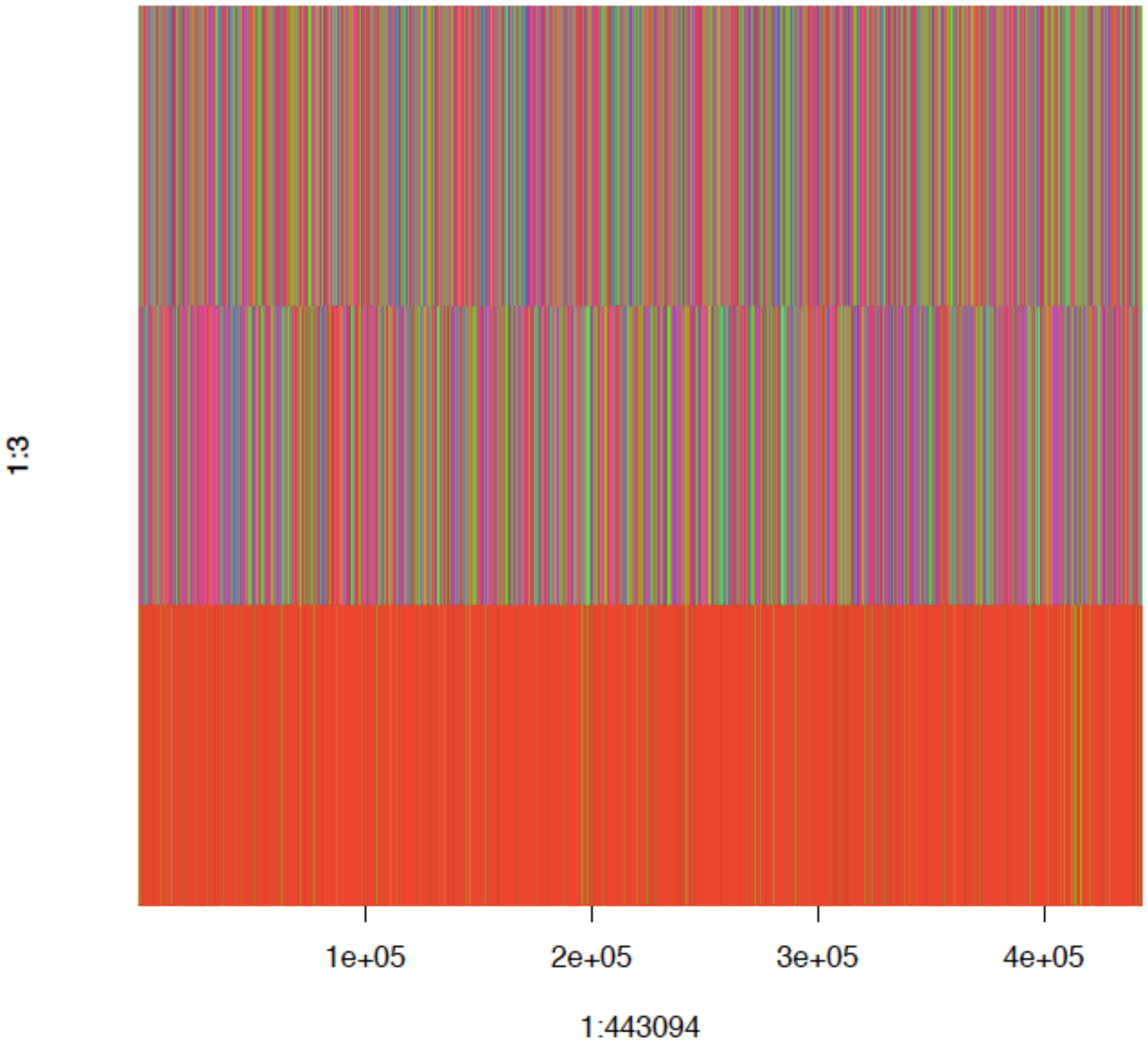}
\caption{Spectroscopic redshifts. p-Adic display, for p = 7.}
\label{fig8}
\end{center}
\end{figure}

The outcomes are shown in the following figures.  Since some of the redshift data
values are just three digits in precision, that was the extent of data precision 
that was used.  See Figures \ref{fig5}, 
\ref{fig8}.  


\begin{figure}
\begin{center}
\includegraphics[width=8cm]{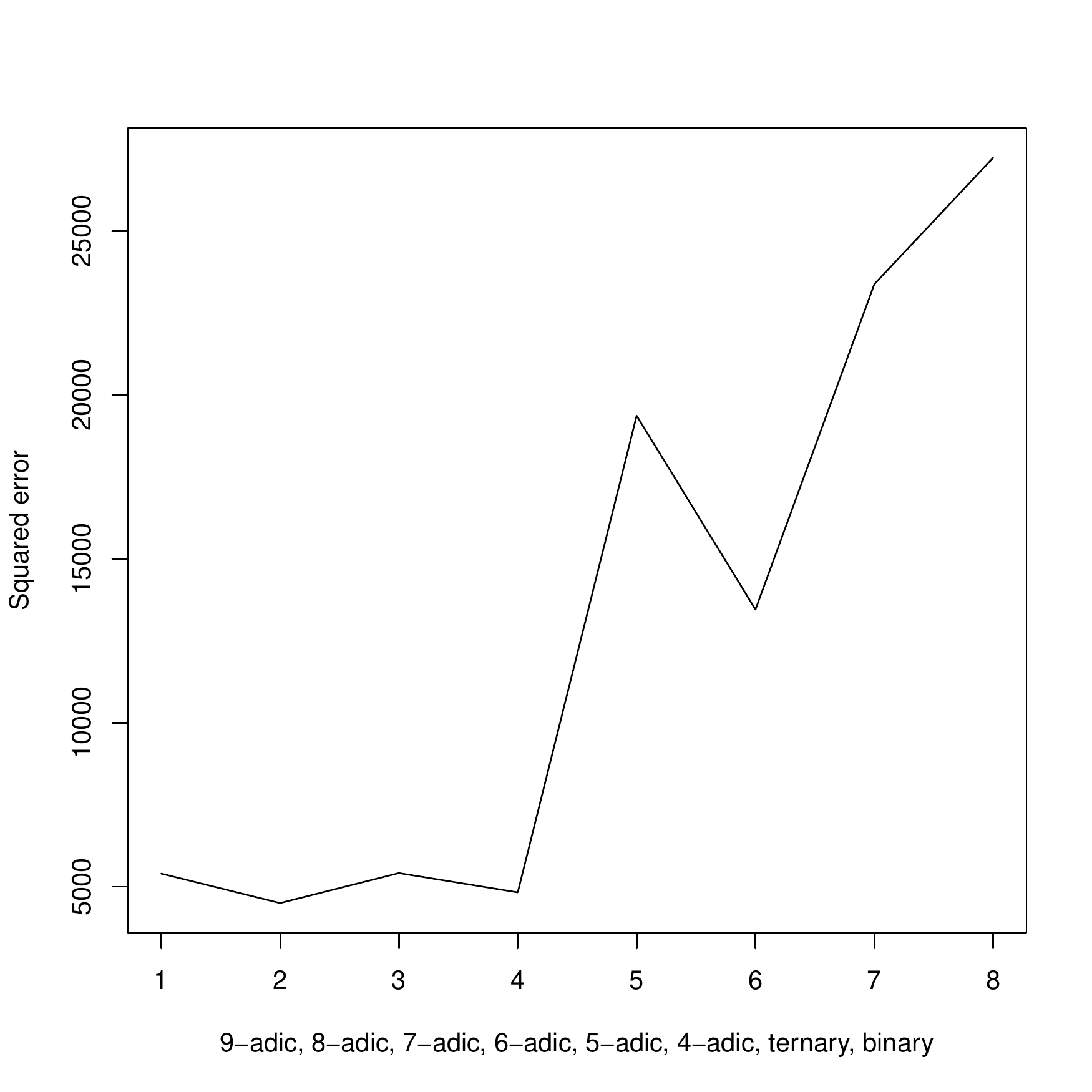}
\caption{Spectroscopic redshifts. Squared distance, i.e.\ error, original 10-adic 
representation, and the sequence of m-adic best fits.}
\label{fig14}
\end{center}
\end{figure}

In Figure \ref{fig14} there is a plot of distances squared, i.e.\ squared
error, between the original (m-adic, with m = 10) data and the other m-adic 
representations.  Note that appropriate normalization (rescaling to $(0,1)$ 
for each m-adic representation) precedes the calculation of distance squared.  

A minor remark follows on Figure \ref{fig14}: the best p-adic fit to our data is 
the 7-adic representation.  

\subsection{m-Adic Fitting of Photometric Redshifts}

As described in the previous subsection, for spectroscopic redshifts, we now 
study the photometric redshifts. 

\begin{figure}
\begin{center}
\includegraphics[clip,trim={0 4cm 0 4cm},width=8cm]{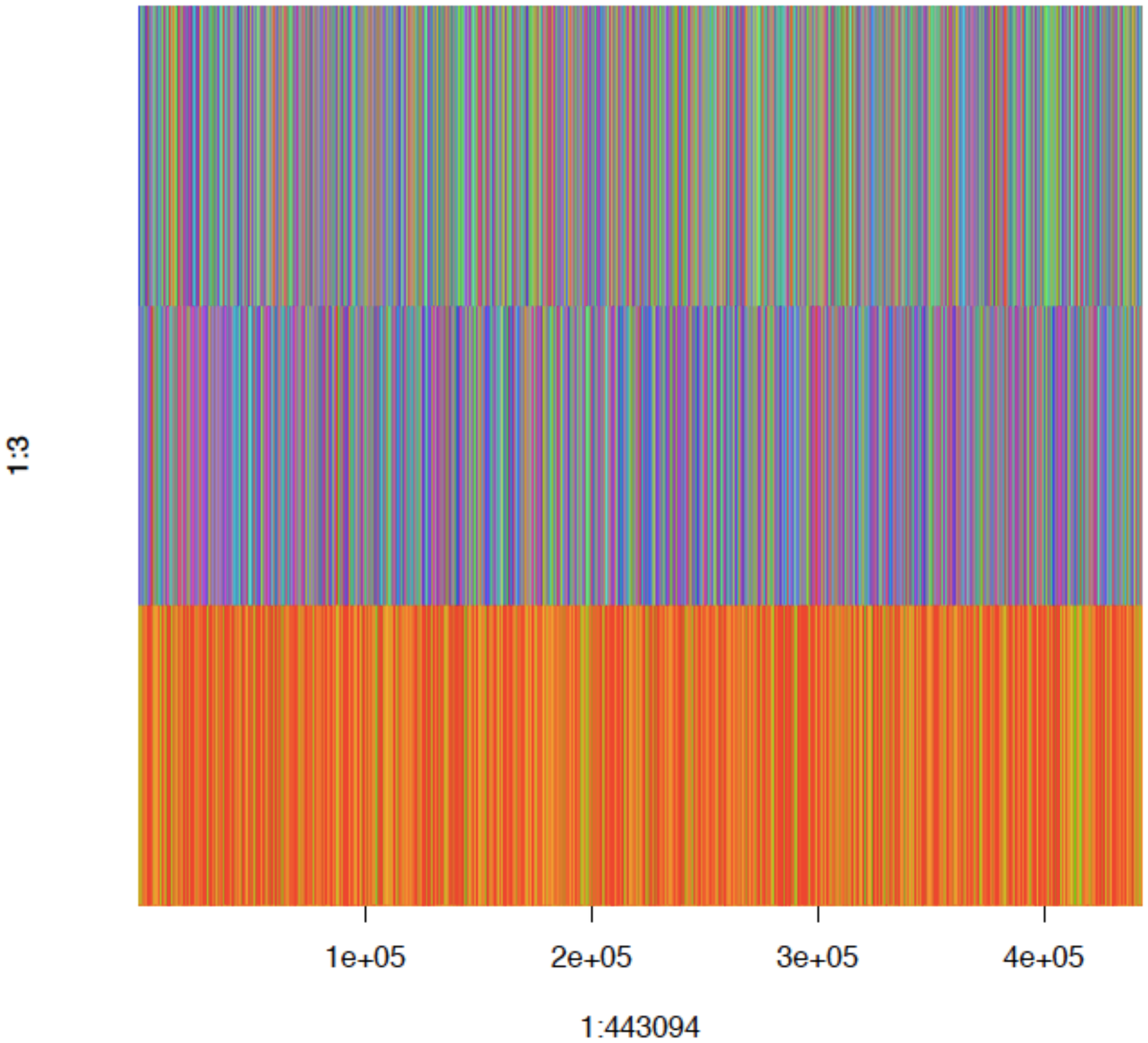}
\caption{Photometric redshifts. 
Initial m-adic display, for m = 10.  Three digits of precision used.}
\label{fig15}
\end{center}
\end{figure}

The outcomes are shown in the following figures.  Three-digit data precision was 
used.  See Figure \ref{fig15}.

\begin{figure}
\begin{center}
\includegraphics[width=8cm]{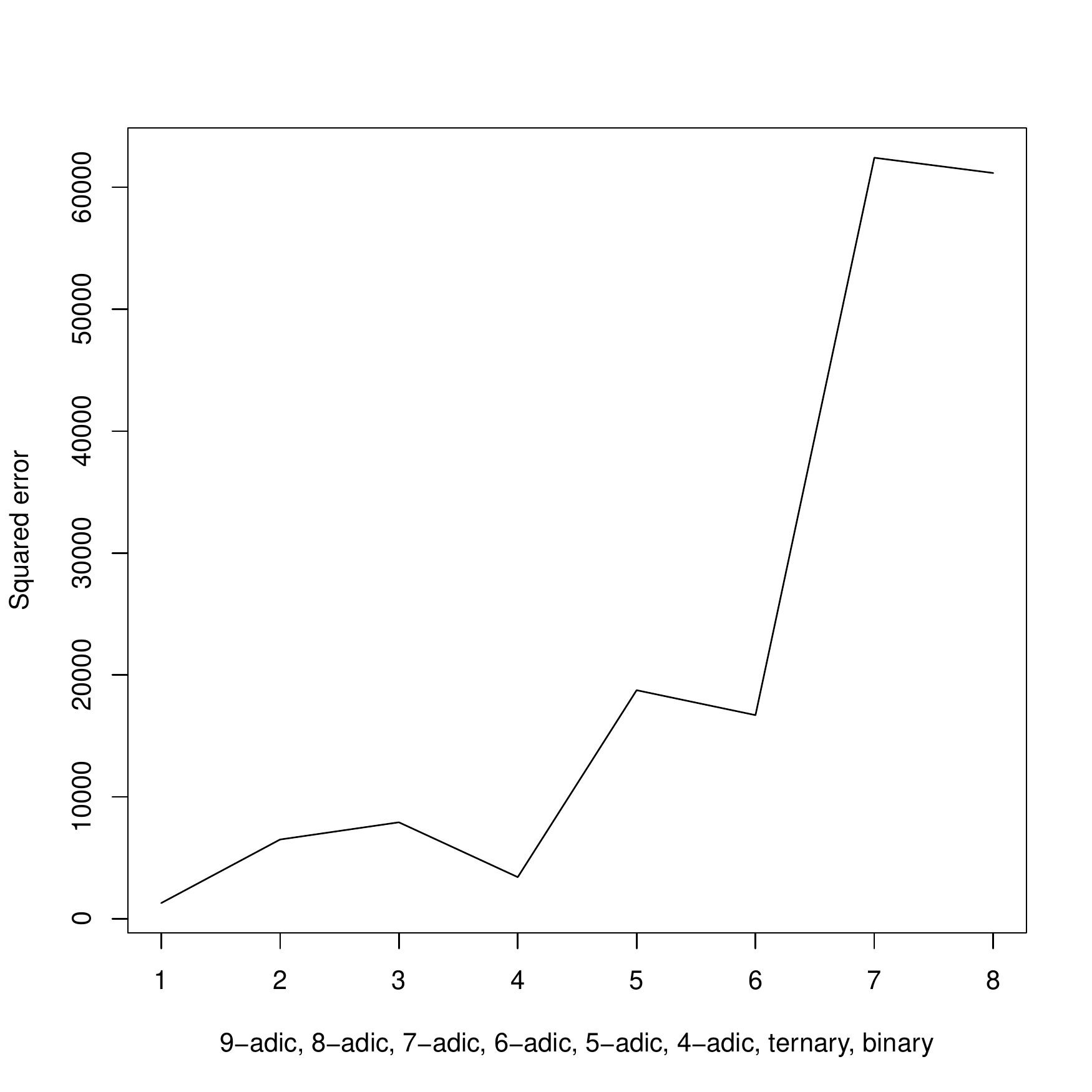}
\caption{Photometric redshifts. Squared distance, i.e.\ error, original 10-adic
representation, and the sequence of m-adic best fits.}
\label{fig24}
\end{center}
\end{figure}

In Figure \ref{fig24} there is a plot of distances squared, between the 
original m-adic representation, for m = 10, and the succession of best fits.  

\subsection{m-Adic Regression: Best Fit of Spectrometric Redshifts by Photometric
Redshifts}

From Table \ref{tab1}, we see that the binary representation, of spectroscopic and 
photometric redshifts, gives the best, closest correspondence.  This is for all 
digits. Since we desire exact matching, as far as possible, rather than just such a 
real-valued degree of approximation, we look further for that objective. 

\begin{table}
\begin{center}
\caption{Totalled distance between spectroscopic and photometric redshifts}
\begin{tabular}{lr} \hline
Representation    &  Distance \\ \hline
Original, m-adic  &  3497.347 \\
m-adic, m = 9     &  3219.545 \\
m-adic, m = 8     &  2960.628 \\
p-adic, p = 7     &  2463.237 \\
m-adic, m = 6     &  2102.937 \\
p-adic, p = 5     &  1798.283 \\
m-adic, m = 4     &   1401.31 \\ 
p-adic, p = 3     &  1009.443 \\
p-adic, p = 2     &  940.8114 \\ \hline
\end{tabular}
\label{tab1}
\end{center}
\end{table}

In Table \ref{tab2}, it is seen that up to 57\% of the digits in the ternary, or 3-adic, 
representations of spectroscopic and photometric redshifts, are identical.  Let us 
see next, if this can be improved upon.  

\begin{table}
\begin{center}
\caption{Identical digits between spectroscopic and photometric redshifts, 
the total number, and as the fraction of all digits in these 443094 objects.}
\begin{tabular}{lrr} \hline
Representation  &  No. identical digits &  Fraction \\  \hline
Original, m-adic  & 508376 & 0.3824441 \\
m-adic, m = 9  & 361332 & 0.2718249 \\
m-adic, m = 8  & 404957 & 0.3046434 \\
p-adic, p = 7 & 446470 & 0.3358731 \\
m-adic, m = 6 & 487841 & 0.3669959 \\
p-adic, p = 5 & 712084 & 0.5356907 \\
m-adic, m = 4 & 745784 & 0.5610427 \\
p-adic, p = 3 & 757357 & 0.5697489 \\
p-adic, p = 2  & 736578 & 0.5541172 \\ \hline
\end{tabular}
\label{tab2}
\end{center}
\end{table}

Table \ref{tab3} displays the most successful outcome here.  We are using just the
first digit of the representation of redshifts, in m-adic representations, for m = 
10, 9, $\dots$, 3, 2.  We find that for the cases of either p-adic with p = 5, 
5-adic, or m-adic with m = 4, we have 98\% identity between spectroscopic and 
photometric redshifts. Thus, from this data set, comprising (SDSS, DR5, Stripe 82)
redshifts for 443094 objects, the desired equivalence between spectroscopic and
photometric redshifts, points to desirability of either 4-adic or 5-adic redshift
encoding.  These, respectively, comprise their values using the digit sets, 
0, 1, 2, 3 and 0, 1, 2, 3, 4.  Most of all in such representations, there are 
natural, implicit hierarchical data representations, i.e.\ here, regular 4-way and 
5-way trees.  If we were to accept a little less identity between the redshifts, 
to have just over 89\% measurement identity, then we would be content with 
either of the ternary (p-adic with p = 3), or binary (p-adic with p = 2) representations.

\begin{table}
\begin{center}
\caption{Compared to Table \ref{tab2}, here just the first digit of precision is used.
Identical digits between spectroscopic and photometric redshifts,
the total number, and as the fraction of all digits in these 443094 objects.}
\begin{tabular}{lrr} \hline
Representation    &  No. identical digits &  Fraction  \\ \hline
Original, m-adic & 366907 & 0.8280568 \\
m-adic, m = 9 & 213872 & 0.4826786 \\
m-adic, m = 8 & 247360 & 0.5582563 \\
p-adic, p = 7 & 262474 & 0.5923664 \\
m-adic, m = 6 & 262474 & 0.5923664 \\
p-adic, p = 5 & 434736 & 0.9811372 \\
m-adic, m = 4 & 434736 & 0.9811372 \\
p-adic, p = 3 & 395490 & 0.8925646 \\
p-adic, p = 2 & 395490 & 0.8925646 \\ \hline
\end{tabular}
\label{tab3}
\end{center}
\end{table}

\section{Conclusions}

It is acknowledged that the experimental work here has only used a training set in order to specify a 
model for matching based on number representation.  If used for regression, it is seen to be a method
of cluster-wise regression, generalizing nearest neighbour regression.  The clustering or binning of the 
data values has a central role in this methodology.  

Precision of measurement is a statistical issue, that was so fundamental to the seminal work of 
Carl Friedrich Gauss.  In this work, our focus has been on clustering, or binning, or interval 
specification. For real valued data, this is an approach to replacing a range of values, in an interval,
with a cluster label, e.g.\ $x + \epsilon \longrightarrow c$ with $x, \epsilon \in \R, c \in \Z_+$.  

Longer term, our objective is more to do with tracking, and in a non-statistical sense, inferring 
structure from the data.  Such structure includes relative distance from the observer, and associated 
with this, inter- and intra-distances for clustered objects.  Central to this is the topology rather 
than geometry manifested by observed (spatial, shaped, ordered) data.   Another, longer term goal, 
is the explicit incorporation of the time dimension.

Extending this methodology is additionally of interest, as written in 
\cite{nbody}, ``Clearly those 
interested in (re)structuring data for any purpose ought to 
keep a close watch for innovative and interesting approaches in the cosmological simulation field in the 
future!''.  This continues: ``However, the structuring of particles with a view towards force calculations
has also something to learn from experience in the cluster analysis field.'' 

Finally, of note, is the potential to relate this work (in a manner, to be determined) with the p-adic 
and adelic number theoretical explanations of dark matter and dark energy, 
\cite{branko}. In \cite{murtagh2017}, there 
is some further discussion of this.  
In \cite{branko}, a section heading is as follows 
(p.\ 40): ``Adelic        
Universe with Real and p-Adic Worlds'', and there is this motivation (p.\ 26):
``Let us use terms real and p-adic to denote those aspects of the universe which
can be naturally described by real and p-adic numbers, respectively.  We conjecture
here that the visible and dark sides of the universe are real and p-adic ones,
respectively.''

In the different context of social science, 
\cite{murtaghspagat} considers hierarchical
representation of extent and degree of change in multivariate time series.  The aim 
is to uncover relationships between social violence and market forces.  There also, 
internal structure arising from inherent heterogeneity in our data is directly used.  

\bigskip

\noindent
{\bf Acknowledgement}

\noindent
The SDSS data were provided by Raffaele D'Abrusco and Giuseppe Longo.

\end{document}